# Qubit Coupling to Reservoir Modes: Engineering the Circuitry to Enhance the Coherence Time

Ahmad Salmanogli
Çankaya University, Engineering faculty, Electrical and Electronic Department, Ankara, Turkey

**Abstract**
In this study, a circuitry model of the coupling of a qubit to reservoir modes is defined to clearly determine the effect of the reservoir modes on the qubit decay and dephasing rates. The main goal is to theoretically calculate the dephasing and decay rate of a qubit, particularly due to the circuitry effect. Firstly, the Hamiltonian of the system (coupling of a qubit to the reservoir modes) is defined and used to derive the time evolution of the density matrix for the qubit energy levels. By calculating the qubit's level density evolutions, one can estimate at which frequency the maximum coupling occurs and, in addition, knows about the role of the electromagnetic bias in the qubit. Secondly, the qubit decay rate is theoretically derived. The results show that the decay rate is strongly affected by circuitry elements such as the qubit capacitor and, more importantly, the coupling capacitor between the qubit and the reservoir modes. As the main result, it is shown that the slight decrease in the coupling capacitors significantly affects the relaxation time even more than the qubit capacitor. Consequently, the dephasing rate, which is the effect of the reservoir modes on the transition frequency of the qubit, is theoretically examined using the Heisenberg-Langevin equation. Finally, by transforming the Heisenberg-Langevin equations into the Fourier domain, the number of photons of the qubit due to coupling to reservoir modes are calculated. This is considered to be the photons generated in the qubit owing to the noise effect. This term significantly influences the qubit coherence time.

**Introduction**
In recent years, one of the critical challenges of quantum computers and quantum processor designers has focused on the decoherence time, the factors affecting it, the sources of decoherence, and its controlling and limiting [1-10]. The decoherence time is simply understood through measurements in the quantum system [1]. The challenge in the solid-state implementation of a quantum processor is how it is possible to create a very long coherence time for qubits in quantum systems [3], by which the quantum properties of the quantum system are strongly preserved. However, it is crucial to know exactly about the decoherence sources and find methods to limit and confine it [4]. An interesting method that has been applied to limit coherence time is to bias the qubit at an optimum point by which the coherence time is significantly increased [4]. Generally, the coherence of a qubit is strongly confined by relaxation from the excited state of the qubit, and also the dephasing phenomenon plays a critical role [2, 5].

The relaxation time originates from the interaction between the qubit and the electromagnetic cavity. The established quantum system must interact (e.g., interaction of the qubit with wires and electrical circuits) because it makes control and readout possible [2]. The value of relaxation time varies from sample to sample and strongly depends on factors such as material fabrication and circuit design [2, 4, 6]. The relaxation time is usually confined by spontaneous emission, the Purcell effect, dielectric loss, quasiparticle tunneling, and flux coupling [8]; however, it is significantly limited by the spontaneous emission arising from the circuitry effect. In other words, the electromagnetic modes generated by the circuit connected to the qubit can handle and affect spontaneous emission [2, 4, 8]. Spontaneous emission is an unavoidable consequence of qubits and environmental coupling [9, 10]. In other words, the coupling electromagnetic field from the reservoir generates photons in the qubit that dramatically manipulate spontaneous emission. In this study, we theoretically calculate the number of photons generated in the qubit owing to the coupling to the reservoir and consequently show that the number of photons generated is sufficiently high to disturb the state of the qubit. Moreover, the other critical term is dephasing, which is understood in terms of the fluctuation of the qubit transition frequency owing to coupling to the environment [7, 8]. From the literature, it is found that it can be generated because of factors such as charge noise, current noise, and $E_c$ noise [8]. Specifically, the generation of dephasing in a quantum system requires a dispersion regime that is determined by the quality factor of the designed circuit [7]. In the dispersion regime, the photons of an oscillator coupled to a qubit induce qubit frequency shifting. The amount of frequency shifting, which causes dephasing, depends on the coupling between the qubits and oscillators (environment effect) [7].

With knowledge of the points mentioned above, it is found that understanding the mechanism of relaxation and dephasing times becomes important to further improve the qubit coherence time [2, 3]. This requires an exact model of the contribution to relaxation and dephasing times. There are some practical methods, such as heterodyne and homodyne, for measuring the spontaneous emission and dephasing of a qubit [1]. In addition, some other approximation approaches, such as single-mode approximation, have been applied to measure decoherence times [2, 3]. However, there is a need for a strong theoretical method to precisely define the change in the coherence time owing to spontaneous emission in a qubit coupled to an environment. In this study, the focus is on the circuitry effect coupled to a qubit by which the spontaneous emission and dephasing times are manipulated. We attempt to define exactly how and by which quantities the coherence time of a qubit is changed and enhanced. Following, the theory and backgrounds of the system are presented.

## Theory and Backgrounds

A schematic of the system, in which N distinct oscillators are coupled to a qubit in the K-space, is illustrated in Fig. 1. For simplicity, it is assumed that the transmission line that connects the qubit to the environment is lossless; nonetheless, the effect of photon loss (system decay rate $\kappa$) will be applied in the Heisenberg-Langevin equation. The schematic shows that N different modes in the reservoir are coupled to the qubit, which causes decoherence in the qubit. The total decoherence containing the relaxation and dephasing factors is $\gamma_c = \sum_{k=1} \gamma_k$, where $\gamma_k$ is the decoherence owing to each oscillator in the K-space.

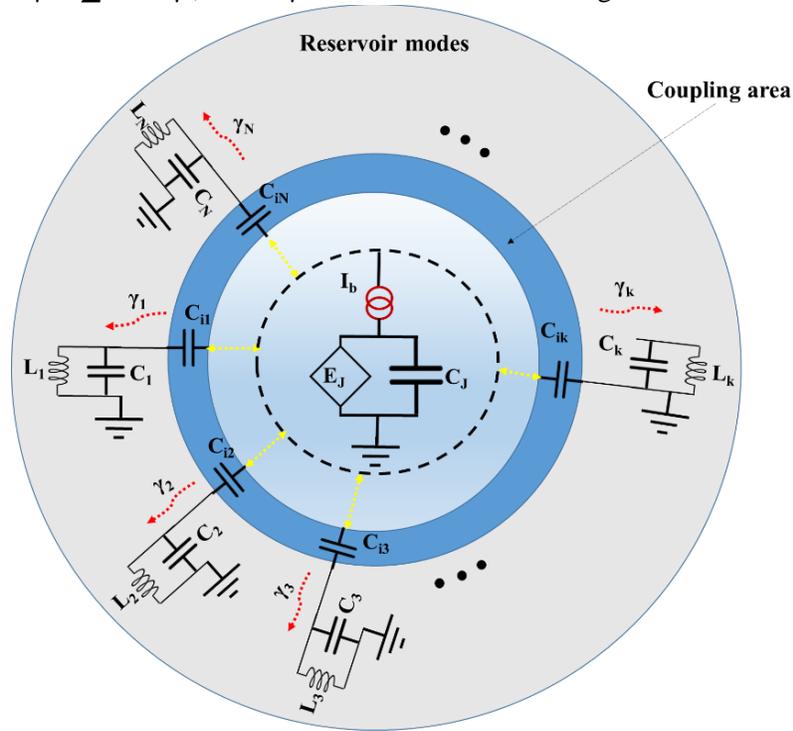

Fig. 1 Schematic of a qubit coupling to the reservoir modes in K-space

The system illustrated in Fig. 1 contains N oscillators capacitively coupling to a qubit. Thus, the Lagrangian of the qubit coupled to the reservoir modes is expressed as:

$$L = \frac{C_j}{2}\dot{\varphi}^2 - E_j \cos\varphi + \sum_{k=1}^{N}\left\{\frac{C_{ik}}{2}\left(\dot{\varphi}-\dot{\varphi}_k\right)^2 + \frac{C_k}{2}\dot{\varphi}_k^2 - \frac{1}{2L_k}\varphi_k^2\right\} \quad (1)$$

where $C_j$, $C_{ik}$, $C_k$, and $L_k$ are qubit capacitor, coupling capacitor between qubit and reservoir, oscillators capacitor and inductor, respectively. In addition, $\varphi$ and $\varphi_k$, are qubit and reservoir oscillators relating node flux considered as the coordinates. To derive the Hamiltonian of the system using the Lagrangian expressed in Eq. 1, one needs to use Legendre transformation to calculate the momentum conjugate Q and $Q_k$ by which the relationship between coordinate (loop flux) and momentum conjugate (node charge) becomes:

$$\begin{pmatrix} Q \\ Q_k \end{pmatrix} = \begin{pmatrix} C_j + \sum_{k=1}^{N} C_{jk} & -\sum_{k=1}^{N} C_{jk} \\ -\sum_{k=1}^{N} C_{jk} & \sum_{k=1}^{N}(C_{jk} + C_k) \end{pmatrix} \begin{pmatrix} \dot{\varphi} \\ \dot{\varphi}_k \end{pmatrix} \quad (2)$$

Thus, the Hamiltonian of the system is given by:

$$H = \frac{1}{2C_{q0}}Q^2 - E_j\cos\varphi + \sum_{k=1}^{N}\left\{\frac{1}{2C_{q1}}Q_k^2 + \frac{1}{2C_{q2}}Q_kQ + \frac{1}{2L_k}\varphi_k^2\right\} \quad (3)$$

where $C_{q0} = C^2/\sum_{k=1:N}(C_{ik}+C_k)$, $C_{q1} = C^2/\{C_j + \sum_{k=1:N}(C_{ik})$, $C_{q2} = C^2/2\sum_{k=1:N}(C_{ik})$, and $C^2 = C_j*\sum_{k=1:N}(C_{ik}+C_k) + \sum_{k=1:N}(C_{ik}*C_k)$. Eq. 3 clearly shows the coupling of the qubit to the reservoir modes through $C_{q2}$, and in addition it is shown that coupling to reservoir modes causes a decrease in the charging energy of the qubit since $C_{q0}$ is greater than $C_j$. In the following, for simplicity the symbol $\sum$ is ignored from the equations and some definitions such as $\sum_{k=1:N}(C_{ik}) \equiv C_{i\Sigma k}$ and $\sum_{k=1:N}(C_k) \equiv C_{\Sigma k}$ are expressed. If one supposes the qubit as an artificial two-level atom and also the reservoir modes as an electromagnetic radiator, thus the Hamiltonian of the system can be considered as the interaction of the electromagnetic field with two-level energy and is presented as:

$$H = \frac{\hbar\omega_q}{2}\sigma^z - E_j\left(\sigma^+ + \sigma^-\right) + \hbar\omega_k b_k^+ b_k + \hbar g_k\left(\sigma^+ b_k + b_k^+\sigma^-\right) \quad (4)$$

where $\sigma^z = \sigma^{11} - \sigma^{00}$, $\sigma^+$, and $\sigma^-$ are the Pauli matrices. In this equation, $\omega_q$ and $\omega_k$ are the resonance frequencies of qubit and reservoir oscillator's frequencies. In addition, $g_k = \{2eC_{i\Sigma k}/(\hbar C^2)\}\sqrt{(\hbar/2Z_k)}$ is the coupling rate between qubit and the reservoir modes, in which $e$ and $Z_k = \sqrt{(L_k/C_{q1})}$ are the electron charge and reservoir oscillator's impedance, respectively.

Following, the study mainly concentrates on the theoretically calculation of the decoherence of the qubit due to the coupling to reservoir modes. We specifically derive the spontaneous emission which is the original factor of the relaxation time and also using the Heisenberg-Langevin equation, the qubit dephasing time will be calculated. For this, the study starts with the calculation of the density matrix elements time evolution by which one can estimate how and by which factor the energy is transferred between different levels of the qubit. To do this, the Hamiltonian expressed in Eq. 4 is divided into free evolution ($H_0$) and interaction ($H_I$) Hamiltonian as:

$$H_0 = \frac{\hbar\omega_k}{2}\sigma^z + \hbar\omega_k b_k^+ b_k$$

$$H_I = \frac{\hbar\Delta\omega}{2}\sigma^z - E_j\left(\sigma^+ + \sigma^-\right) + \hbar g_k\left(\sigma^+ b_k + b_k^+\sigma^-\right) \quad (5)$$

where $\Delta\omega = \omega_q - \omega_k$. The density operator at later time is $\rho(t) = U_1(t)U_2(t)\rho(0)U_2(t)^+U_1(t)^+$, where $U_1(t) = \exp(-jH_0t/\hbar)$ and $U_2(t)=\exp(-jH_It/\hbar)$ [11] are the time development operators. We assume that the qubit is in the upper state and the reservoir modes be in the vacuum state; thus, the density operator at later time becomes:

$$\rho(t) = U_1(t)U_2(t)|0> \begin{bmatrix} 1 & 0 \\ 0 & 0 \end{bmatrix} <0|U_2(t)^+ U_1(t)^+ \longrightarrow = \begin{bmatrix} \rho_{11}(t) & \rho_{12}(t) \\ \rho_{21}(t) & \rho_{22}(t) \end{bmatrix}$$

$$\begin{cases} \rho_{11}(t) = \left\{ \cos^2 t\sqrt{\Delta_\alpha^2 + g_k^2} + \frac{\Delta\omega^2}{4} \frac{\sin^2 t\sqrt{\Delta_\alpha^2 + g_k^2}}{\Delta_\alpha^2 + g_k^2} \right\} |0><0| \\ \\ \rho_{12}(t) = \left\{ j\left(-\frac{E_j}{\hbar}\right) \cos t\sqrt{\Delta_\alpha^2 + g_k^2} \frac{\sin t\sqrt{\Delta_\alpha^2 + g_k^2}}{\sqrt{\Delta_\alpha^2 + g_k^2}} \right\} |0><0| \\ \\ \rho_{22}(t) = \left\{ \left(\left(\frac{E_j}{\hbar}\right)^2 + g_k^2 n^2\right) \frac{\sin^2 t\sqrt{\Delta_\alpha^2 + g_{k0}^2}}{\Delta_\alpha^2 + g_k^2} \right\} |0><0| \end{cases} \quad (6)$$

where $\Delta_\alpha^2 = \Delta\omega^2/4 + (E_j/\hbar)^2 + g_k^2 n_q^2$. In this expression $n_q$ defines the number of photons in qubit generated due to the interaction with the reservoir and finally we will show that this term significantly affects the qubit coherence time. To make an estimation about $n_q$, one can use the total Hamiltonian of the system to create the Heisenberg-Langevin equations [11-15] given by:

$$\dot{a} = -\left(j\omega_q + \frac{\kappa}{2}\right)a - jg_k\left(b_k - b_k^+\right) + 2\kappa a_{in}$$

$$\dot{b}_k = -j\omega_k b_k - jg_k\left(a - a^+\right) \quad (7)$$

where $\kappa$ and $a_{in}$ are the photons loss in qubit due to the interaction with environment and thermal noise effect, respectively. In this equation, for simplicity the nonlinear terms due to the kinetic energy of the qubit ($E_j$) is ignored. With transformation of Eq. 7 into the Fourier domain [12, 13]:

$$\left(j(\omega_q + \omega) + \frac{\kappa}{2}\right)a = -jg_k\left(b_k - b_k^+\right) + 2\kappa a_{in} \longrightarrow a = -\frac{jg_k}{\left(j(\omega_q + \omega) + \frac{\kappa}{2}\right)}\left(b_k - b_k^+\right) + \frac{2\kappa}{\left(j(\omega_q + \omega) + \frac{\kappa}{2}\right)} a_{in}$$

$$j(\omega_k + \omega)b_k = -jg_k\left(a - a^+\right) \longrightarrow b_k = -\frac{g_k}{(\omega_k + \omega)}\left(a - a^+\right) \quad (8)$$

where $\omega$ is the sweeping frequency and in the simulations, we suppose it can change around $\omega_q$. Using Eq. 8, one can calculate $n_q = <a^+a>$, $n_k = <b_k^+b_k>$, and $n_{in} = <a_{in}^+a_{in}>$ indicating the qubit and reservoir modes number of photons, and thermal photon number, respectively. The expression related to the number of photons are theoretically derived and expressed in Appendix A, from which the number of photons of the qubit due to the coupling effect is calculated as:

$$n_q = \frac{1}{\left((\omega_q + \omega)^2 + \frac{\kappa^2}{4}\right) - \frac{4g_{k0}^4}{(\omega_k + \omega)^2}} \left\{ g_{k0}^2 + 4\kappa^2 n_{in} + \frac{2g_{k0}^4}{(\omega_k + \omega)^2} \right\} \quad (9)$$

The equation shows that the number of photons of qubit due to the coupling to the reservoir is strongly affected by the coupling rate. Thus, the coupling rate between qubit and reservoir modes significantly determines the number of photons created due to the noise effect. In addition, Eq. 8 clearly demonstrates that the phase sensitive cross-correlation photon number between the qubit and reservoir modes equals zero

<abk> = 0. Using this, one can find that the entanglement metric between two modes always equals to zero $\varepsilon = <ab_k>/\sqrt{(<a^+a><b_k^+b_k>)}$ [12, 13]. Thus, as a clear point, it is shown that there are no any correlations between the qubit and reservoir modes photons even at $\omega_q = \omega_k$.

The simulation results relating to the number of photons, $n_q$ and $n_k$ are illustrated in Fig. 2. In this simulation, $n_q$ and $n_k$ are generated owing to the coupling of the reservoir to the qubit, which means that the noise is responsible for adding some photons to the system. Because the quantum system works with a low number of photons, if the number of photons coupled to the qubit becomes comparable because of the noise, it dramatically affects the qubit quantum properties. For instance, if the qubit is initially prepared to operate in vacuum state (e.g., at temperatures less than 50 mK, the thermal occupation is less than 0.005 photons at 5.64 GHz and this state is considered the vacuum state [10]), increasing the number of photons disturbs the vacuum state and shifts the state to other states. Fig. 2a shows the change in $n_q$ and $n_k$ with respect to $\omega_k$ as the coupling rate between the qubit and reservoir modes is limited to $0.1g_k$. It is shown that, at $\omega_q = \omega_k$ the number of photons generated in the qubit and reservoir modes becomes equal to $n_q = n_k$. One of the interesting results of this study is demonstrated in Fig. 2b, in which increasing $C_j$ causes the decrease of the number of photons generated owing to coupling to reservoir modes. This means that increasing $C_j$ confines the noise effect by which photons can be induced into the qubit from the reservoir modes. One of the main conclusions of this study is that the latter phenomenon affects relaxation time. In other words, spontaneous emission is severely controlled by $C_j$, by which the photons generated because of environmental noise are limited. The other simulation results for $n_q$ and $n_k$ in which the effect of the coupling rate $g_k$ is investigated, are shown in Appendix B.

Consequently, the qubit number of photons generated by noise is used to calculate $\Delta\alpha$ to be substituted in Eq. 6 to examine the density matrix elements. For two different numbers of photons of the qubit, such as $n_q = 0.005$ and $n_q = 0.2$, the time variation of the distribution of $\rho_{11}(t)$ and $\rho_{22}(t)$ ($<n|\rho_{11}(t)|n>$ and $<n|\rho_{22}(t)|n>$) is calculated, and the results are shown in Fig. 3. In this simulation, we deliberately ignored the effect of $E_j$ to study only the effect of coupling rate $g_k$. The results indicate that for $n_q = 0.005$, $\rho_{22}(t)$ fluctuates around zero, and the maximum oscillation of $\rho_{11}(t)$ occurs around the detuning frequency $\Delta\omega = 0$ ($\omega_q = \omega_k$). In other words, at approximately $\Delta\omega = 0$, the maximum coupling between the reservoir modes and qubit occurs. In addition, the mentioned effect is intensified when the number of qubit photons is increased. Fig. 3b shows that as the number of qubit photons arising from the noise increases to 0.2, the resonance amplitude increases and reaches the maximum value around $\Delta\omega = 0$. This figure clearly demonstrates the effect of the photons generated owing to the noise in the system, by which the excited state and ground state coupling are dramatically affected. This effect increases the relaxation rate of the system, resulting in a decrease in the coherence time.

In the following, the focus is laid on the relaxation time calculation due to the qubit coupling to the reservoir modes. There are some effects that affect relaxation; however, we only focus on the spontaneous decay rate and Purcell effect. For this, the free evolution part of the Hamiltonian expressed in Eq. 4 is removed and by moving into the interaction picture, the interaction Hamiltonian becomes:

$$H_{inc-p} = \sum_{k=1} \hbar g_k \left( \sigma^+ b_k e^{j(\omega_q - \omega_k)t} + b_k^+ \sigma^- e^{-j(\omega_q - \omega_k)t} \right) \quad (10)$$

For spontaneous decay rate calculation [10], one needs to integrate the equation of motion for density operator in a small interval $\Delta t$: the equation $\partial\rho(t)/\partial t = [H_{inc-p}, \rho(t)]/j\hbar$ can be expressed as $\rho(t+\Delta t) - \rho(t) = 1/j\hbar \int_{0-\Delta t} dt\, [H_{inc-p}, \rho(t)]$. Furthermore, one can add the second order terms as $(1/j\hbar)^2 \int_{0-\Delta t} dt \int_{0-t_1} dt_1 \, [H_{inc-p}, [H_{inc-p}, \rho(t)]]$ to increase the accuracy of the calculation. Calculation of $\rho(t+\Delta t) - \rho(t)$ gives some information about the change of $\rho(t)$ after $\Delta t$, which means that after $\Delta t$ how much the field generated due to the spontaneous emission is coupled to the reservoir fields. In other words, how much the qubit is radiated into the continuum. This is the important issue that the study wants to answer. Since we are only interested in the state of the qubit, not the reservoir, the density operator can be prepared in an effective manner as $\rho_A(t)$

= $Tr_F(\rho(t))$. As we assumed that the field is in vacuum state $\rho_F(t) = |0\rangle\langle 0|$, therefore the equation of motion for the density operator becomes $\rho(t+\Delta t) - \rho(t) = (1/j\hbar)^2 \int_{0-\Delta t} dt \int_{0-t_1} dt_1 Tr_F[H_{inc-p}(t),[H_{inc-p}(t_1), \rho_F(t)*\rho_A(t)]]$. In this equation, it can be easily show that the first term equals to zero [10]. In the present integration $Tr_F[H_{inc-p}(t),[H_{inc-p}(t_1), \rho_F(t)*\rho_A(t)]] = Tr_F[H_{inc-p}(t)H_{inc-p}(t_1)\rho_F(t)*\rho_A(t)] + Tr_F[\rho_F(t)*\rho_A(t) H_{inc-p}(t)H_{inc-p}(t_1)] - Tr_F[H_{inc-p}(t)\rho_F(t)*\rho_A(t) H_{inc-p}(t_1)] - Tr_F[H_{inc-p}(t_1)\rho_F(t)*\rho_A(t) H_{inc-p}(t)]$ [10].

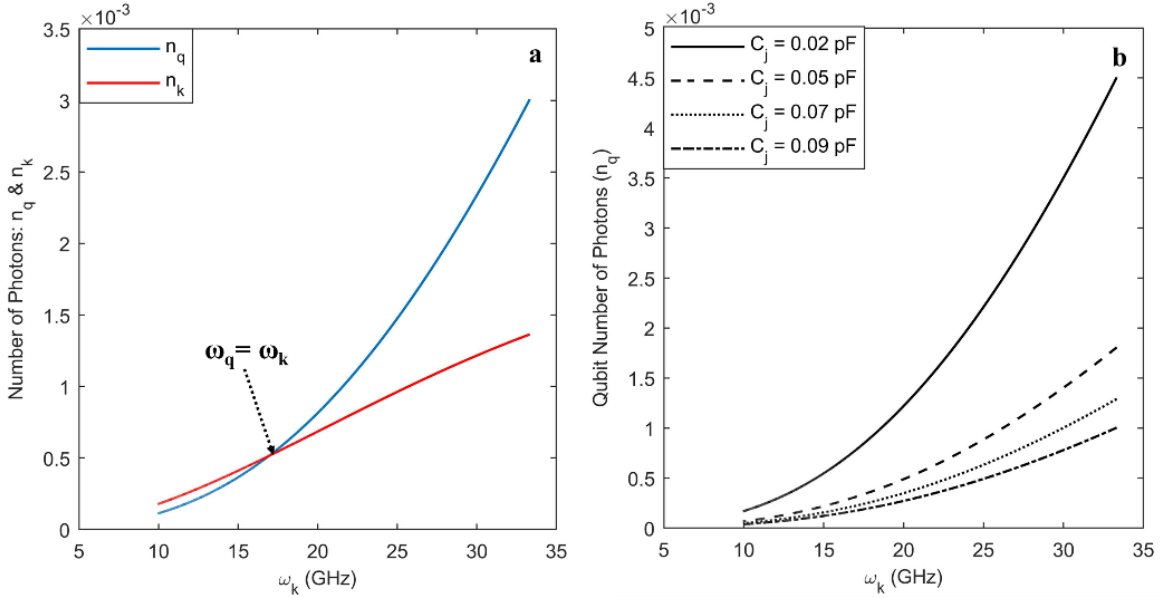

Fig. 2 Effects of the coupling between qubit and reservoir modes on the number of photons of a) qubit and reservoir modes number of photons, b) qubit capacitance effect on number of photons $n_q$ at 10 mK and coupling rate $0.1g_k$; $C_j = 0.03$ pF, $C_{jk} = 0.05$ pF, $L_k = 5$ nH, $C_k$ changes in the range [0.18 pF:2.02 pF].

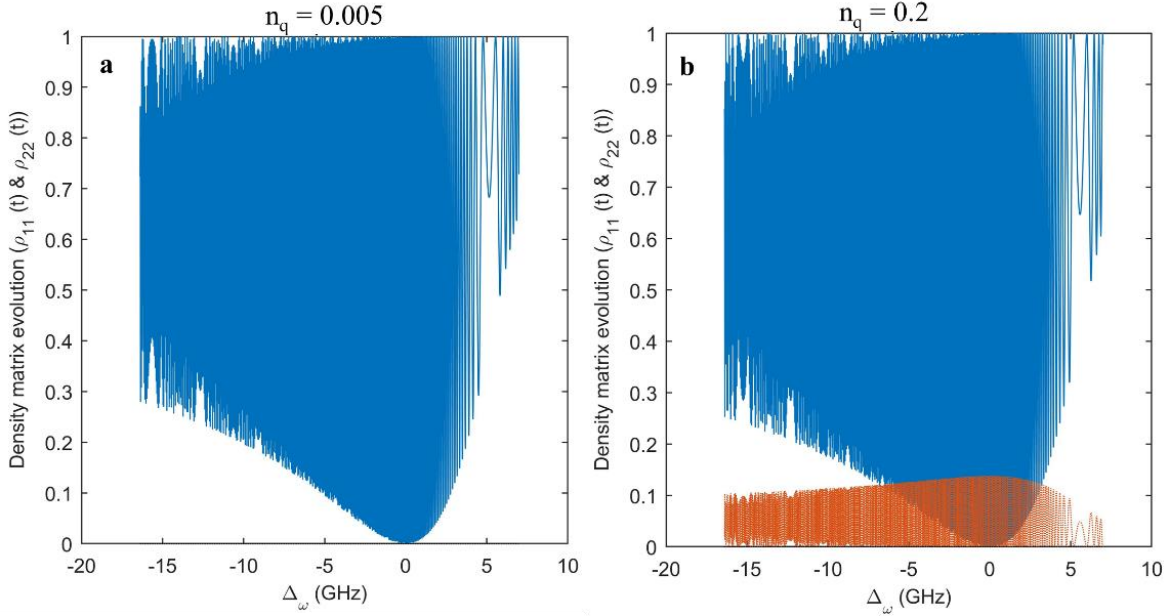

Fig. 3 Time evolution of density matrix elements $\rho_{11}(t)$ (blue) and $\rho_{22}(t)$ (red) vs detuning frequency (GHz) for different qubit number of photons a) $n_q = 0.005$ and b) $n_q = 0.4$; $E_j = 0$; $C_j = 0.03$ pF, $C_{jk} = 0.05$ pF, $L_k = 5$ nH, $C_k$ changes in the range [0.18 pF:2.02 pF].

Consequently, the spontaneous emission factor is derived using:

$$\gamma_s = \int_0^{\Delta t} dt_1 \int_0^{t_1} dt_2 \langle 0| \sum_k \sum_{k'} g_k g_{k'} b_k^+ b_{k'} e^{j(\omega_q - \omega_k)t_1} e^{j(\omega_q - \omega_{k'})t_2} |0\rangle \longrightarrow \int_0^{\Delta t} dt_1 \int_0^{t_1} dt_2 \sum_k |g_k|^2 e^{j(\omega_q - \omega_k)(t_1 - t_2)} \quad (11)$$

In this equation, the double sum in the first part is reduced to a single sum using $b_k^+ b_{k'}|0\rangle = \delta_{kk'}|0\rangle$. To solve the integral in Eq. 11, we initially need to express sum in the form of integral as $\sum_k = \int d\mathbf{k}\, D(k)$, where $D(k)$ is the density of state $(V/4\pi^3)$ and $d\mathbf{k} = k^2 \sin(\theta)\, d\Phi\, d\theta\, dk$ satisfies the integration in a given volume in K-space. In this equation V is the volume, where the modes are quantized. Consequently, by changing summation into integration, Eq. 11 is re-written as:

$$\gamma_s = \int_0^{\Delta t} dt_1 \int_0^{t_1} dt_2 \int_0^{2\pi} d\phi \int_0^{\pi} \sin\theta d\theta \int_0^{\infty} dk\, k^2\, |g_k|^2\, D(k) e^{j(\omega_q - \omega_k)(t_1 - t_2)} \quad (12)$$

where $k = \omega_k/c$ is the wave vector and in addition, c is the speed of light in free space. Solving Eq. 12 leads to calculate the spontaneous emission rate $(\gamma_s/\Delta t)$ presented by:

$$\Gamma_1 = \frac{8\pi^2 e^2}{\hbar c^3} \frac{C_{i\Sigma k}^2 C_{q1}}{C_j^2 (C_{i\Sigma k} + C_{\Sigma k})^2} \omega_q^3 D(k) \quad (13)$$

This is the rate at which the qubit radiates its energy into the continuum or, in an exact expression, the rate of the qubit's emission into the electromagnetic vacuum modes. Because the spontaneous emission rate is a critical factor that limits the coherence time, many designers want to manipulate and limit the rate. Eq. 13 shows that the degree of freedom that designers can specifically focus on, includes $C_j$ (and maybe $C_{jk}$), which is the qubit capacitor (coupling capacitor between qubit and reservoir). As clearly shown in Eq. 13, increasing $C_j$ causes a decrease in the rate, which means that the coherence time increases. For instance, in a transmon qubit, a large capacitor is placed parallel to the qubit to increase $E_j/E_c$ [8, 13]. Using this, Eq. 13 shows that the coherence time is significantly increased in the transmon qubit. As the main conclusion, Eq. 13 suggests that the manipulation of $C_j$ is an effective way to control and limit relaxation time. In addition, the other factor that can be engineered to manipulate the relaxation time is $C_{q1}$ which is a function of $C_j$, $C_{i\Sigma k}$ and $C_{\Sigma k}$. Therefore, one can (maybe) manipulate $C_{i\Sigma k}$ not $C_{\Sigma k}$ by adding an extra degree of freedom to control the spontaneous decay rate through $C_{q1}$. This point is discussed in the last section of this study.

The other important factor in the relaxation time of the qubit is the Purcell effect, in which the spontaneous emission rate is changed. This effect has been calculated using Fermi's golden rule: $\gamma_k = 1/T_k = \kappa g_k^2/\Delta\omega^2$ [8]. Therefore, the relaxation times of the qubit due to coupling to the reservoir modes become $T_s = 1/(\Gamma_1 + \gamma_k)$, where $\Gamma_1 > \gamma_k$.

Finally, to calculate the decoherence time, we need to know about the qubit's dephasing time, which is issued because of the fluctuation of the qubit transition frequency due to coupling to the reservoir modes. In this study, this factor is theoretically derived using Eq. 7. We initially calculated $b_k$ in the steady state, which is equal to $\langle b_k \rangle = g_k(a^+ - a)/\omega_k$. Substituting the steady-state form of $\langle b_k \rangle$ into $da/dt$ in Eq. 7, the modified equation is given by [15]:

$$\dot{a} = -\left(j\left\{\omega_q - \frac{2g_k^2}{\omega_k}\right\} + \frac{\kappa}{2}\right)a + 2\kappa a_{in} \quad (14)$$

The expressions inside the curly brackets in Eq. 14 shows that the transition frequency of the qubit is changed by a factor of $2g_k^2/\omega_k$. The expression calculated here is the dephasing rate ($\gamma_\varphi = 1/T_\varphi$) due to the circuitry effect (coupling of a qubit to N discrete resonators in K-space); however, one can consider other factors, such as the charge noise effect or flux noise on the dephasing time [8]. It is found from Eq. 14 that the dephasing time is directly manipulated by the qubit and reservoir coupling rate, and it also inversely depends on the reservoir resonator frequency.

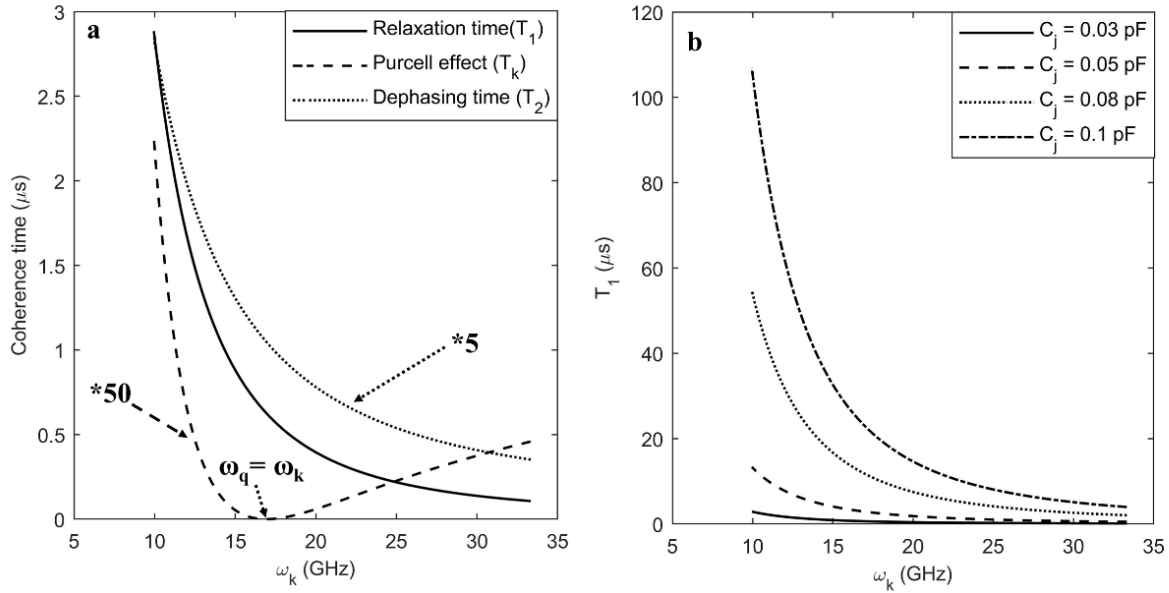

Fig. 4 a) system decoherence time ($T_1$: spontaneous emission time, $T_2$: dephasing time, and $T_k$: relaxation time changing due to Purcell factor) vs. reservoir mode frequency (GHz), b) spontaneous emission time for different qubit capacitors; $C_j = 0.03$ pF, $C_{jk} = 0.05$ pF, $L_k = 5$nH, $C_k$ changes in the range [0.18 pF:2.02 pF].

The simulation results for the effect of the reservoir resonators on the coherence time contributed factors are shown in Fig. 4. In Fig. 4a, for a clear illustration, the values of the dephasing time and Purcell factor are multiplied by 5 and 50, respectively. As expected, the value of the relaxation time is greater than that of the dephasing time, and it decreases as the reservoir frequency increases. This implies that if one needs to design a qubit to operate near the THz frequency, the coupling to the reservoir is strongly increased. In other words, the coherence time is significantly decreased when the qubit resonance frequency is increased. However, as suggested in the literature [1-8], relaxation time is the dominant factor limiting coherence time. In addition, we determined from Eq. 12, the efficient factor that can be manipulated to limit the relaxation time is $C_j$. Fig. 4b shows the effect of $C_j$ on relaxation time. The results demonstrate that increasing $C_j$ effectively confines relaxation time. In other words, this is why the transmon qubit has shown a long coherence time of approximately ms [8].

Finally, the study suggests that another method can be applied to control and limit the reservoir effect, which is the effect of $C_{jk}$ in the system. We know from the system shown in Fig. 1 that $C_k$ cannot be manipulated because $C_k$ is an intrinsic feature of the environment; however, it is possible to manipulate the coupling capacitor effect. For manipulation, one needs to change the connection (transmission line) between the reservoir and qubit. For instance, we usually use a probe to measure the specification of the qubit; thus, we can manipulate the specification of the probe used to connect to the qubit. Because of this fact, the author thinks that it may be possible to even slightly manipulate, or at least affect, the value of $C_{jk}$. Therefore, the last point of this study focuses on this subject. What happens if the designer can change $C_{jk}$? The simulation results are shown in Fig. 5. It is shown that if the value of $C_{jk}$ is decreased by 0.04 pF, the number of photons generated due to the noise from the reservoir applied in the qubit is strongly decreased (Fig. 5a). In addition, Fig. 5b demonstrates that the time variation of the distribution, which shows the coupling between the excited state and ground state, is severely limited very close to $\Delta\omega$. This means that decreasing $C_{jk}$ confines the coupling between the excited state and ground state owing to the noise effects. As a main conclusion, $C_{jk}$ effect is studied on the relaxation time and dephasing time, which are illustrated in Fig. 5c and 5d, respectively. The results show that decreasing slightly in $C_{jk}$ strongly increases the relaxation time from 0.7 μs to 18 μs when $\omega_q = \omega_k$. Similarly, the dephasing time increases as $C_{jk}$ decreases. Consequently, the comparison between Fig. 4b and 5c indicates that the effect of $C_{jk}$ is stronger than that of $C_j$ by which the relaxation time can be manipulated.

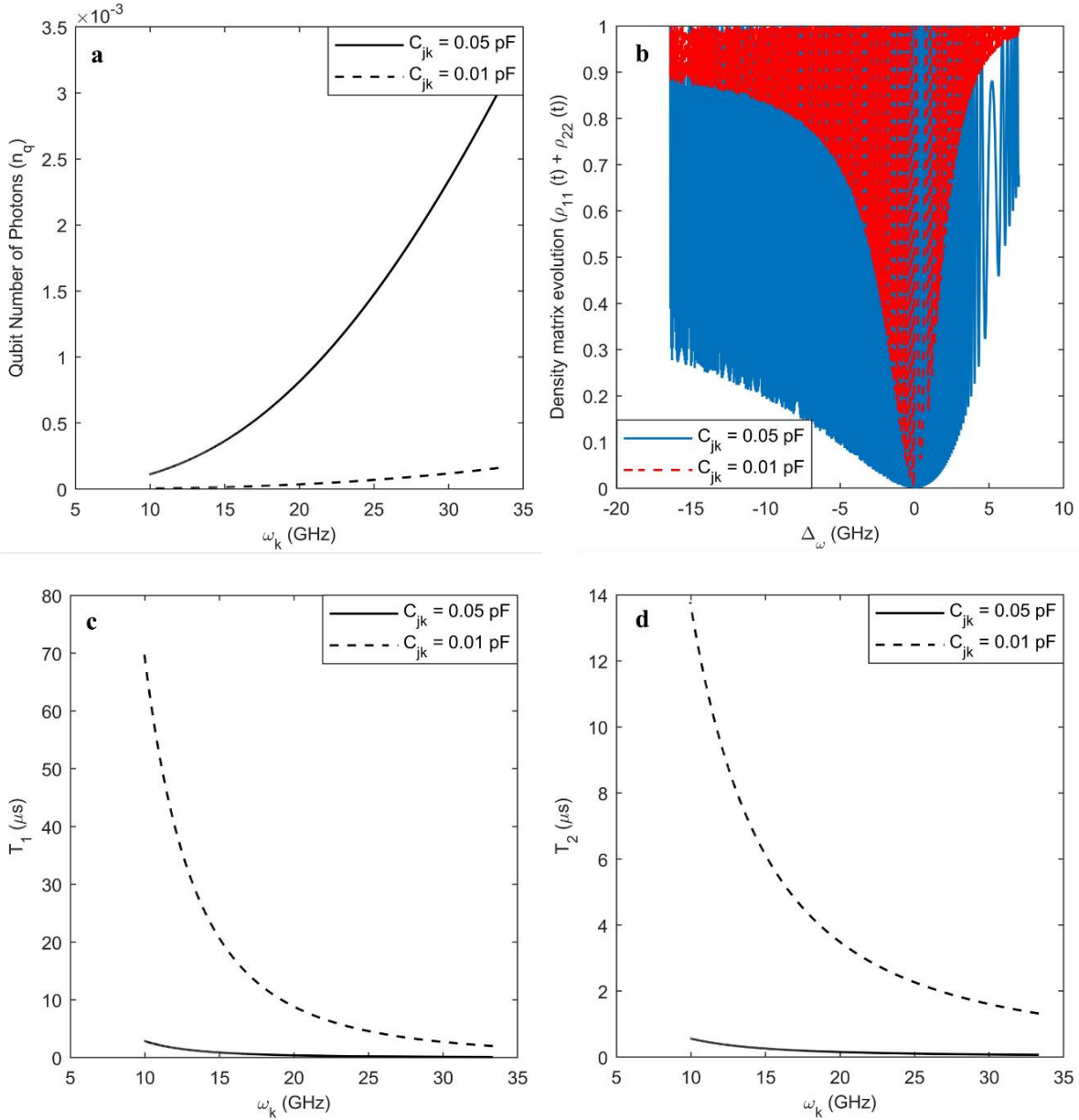

Fig. 5 a) qubit number of photons at $\omega = \omega_q$, coupling rate $0.1g_k$ and T = 10 mK, b) time variation of distribution for $n_q = 0.005$ and $E_j = 0$, c) spontaneous emission time (μs), and d) dephasing time (μs); $C_j$ = 0.03 pF, $L_k$ = 5nH, $C_k$ changes in the range [0.18 pF:2.02 pF].

**Conclusions**

In this study, we defined a system that shows the effect of reservoir modes on a qubit and attempted to theoretically derive the decoherence time due to the circuitry effect. Therefore, the system was defined, and the related Lagrangian and Hamiltonian were calculated. Using the Hamiltonian of the system, the time evolution of the density matrix and number of photons in the qubit and reservoir modes were examined. In the following, we theoretically calculated the dephasing time using the Heisenberg-Langevin equation; consequently, the relaxation time of the system was calculated. From the derived relationship, we find that there are two essential factors for manipulating the coherence time of the qubit. The degrees of freedom introduced by the system are the qubit capacitor and the capacitors that connect the qubit to reservoir modes. The results showed that by engineering the values of these capacitors, the coherence time of the qubit was strongly enhanced. This was mainly attributed to the confinement of the photons generated because of the reservoir modes noise affecting the qubit. In addition, we found that the effect of the coupling capacitor to subside the noise effect is more efficient than of the qubit capacitor to enhance the coherence time.

**Appendix A:**

In this appendix, we tried to calculate the number of photons of qubit and reservoir. From Eq. 8 in main article, a, a⁺, b$_k$, and b$_k$⁺ can be expressed as:

$$a = -\frac{jg_{k0}}{\left(j(\omega_q+\omega)+\frac{\kappa}{2}\right)}(b_k - b_k^+) + \frac{\sqrt{2\kappa}}{\left(j(\omega_q+\omega)+\frac{\kappa}{2}\right)}a_{in}, \quad a^+ = -\frac{-jg_{k0}}{\left(-j(\omega_q+\omega)+\frac{\kappa}{2}\right)}(b_k^+ - b_k) + \frac{\sqrt{2\kappa}}{\left(-j(\omega_q+\omega)+\frac{\kappa}{2}\right)}a_{in}^+ \quad (A1)$$

$$b_k = -\frac{g_{k0}}{(\omega_k+\omega)}(a - a^+), \quad b_k^+ = -\frac{g_{k0}}{(\omega_k+\omega)}(a^+ - a)$$

Using Eq. A1, the related photon numbers are calculated as:

$$n_q = \langle a^+ a \rangle = \frac{g_{k0}^2}{\left((\omega_q+\omega)^2 + \frac{\kappa^2}{4}\right)}\left(2\langle b_k^+ b_k \rangle + 1\right) + \frac{2\kappa}{\left((\omega_q+\omega)^2 + \frac{\kappa^2}{4}\right)}\langle a_{in}^+ a_{in} \rangle \quad (A2)$$

$$n_k = \langle b_k^+ b_k \rangle = \frac{g_{k0}^2}{(\omega_k+\omega)^2}\left(2\langle a^+ a \rangle + 1\right)$$

Finally, one can use the equations in a matrix form to calculate the photons number as:

$$\begin{bmatrix} n_q \\ n_k \end{bmatrix} = \begin{bmatrix} 1 & \frac{-2g_{k0}^2}{\left((\omega_q+\omega)^2 + \frac{\kappa^2}{4}\right)} \\ \frac{-2g_{k0}^2}{(\omega_k+\omega)^2} & 1 \end{bmatrix}^{(-1)} \begin{bmatrix} \frac{g_{k0}^2 + 2\kappa n_{in}}{\left((\omega_q+\omega)^2 + \frac{\kappa^2}{4}\right)} \\ \frac{g_{k0}^2}{(\omega_k+\omega)^2} \end{bmatrix} \quad (A3)$$

**Appendix B:**
In this appendix, the effects of the coupling between qubit and reservoir modes on the number of photons of qubit and reservoir modes are studied. It is shown that increasing coupling rate from $0.1g_k$ to $g_k$ significantly increase the number of photons due to noise in qubit. This effect can dramatically affect the qubit designed initially to operate in vacuum state.

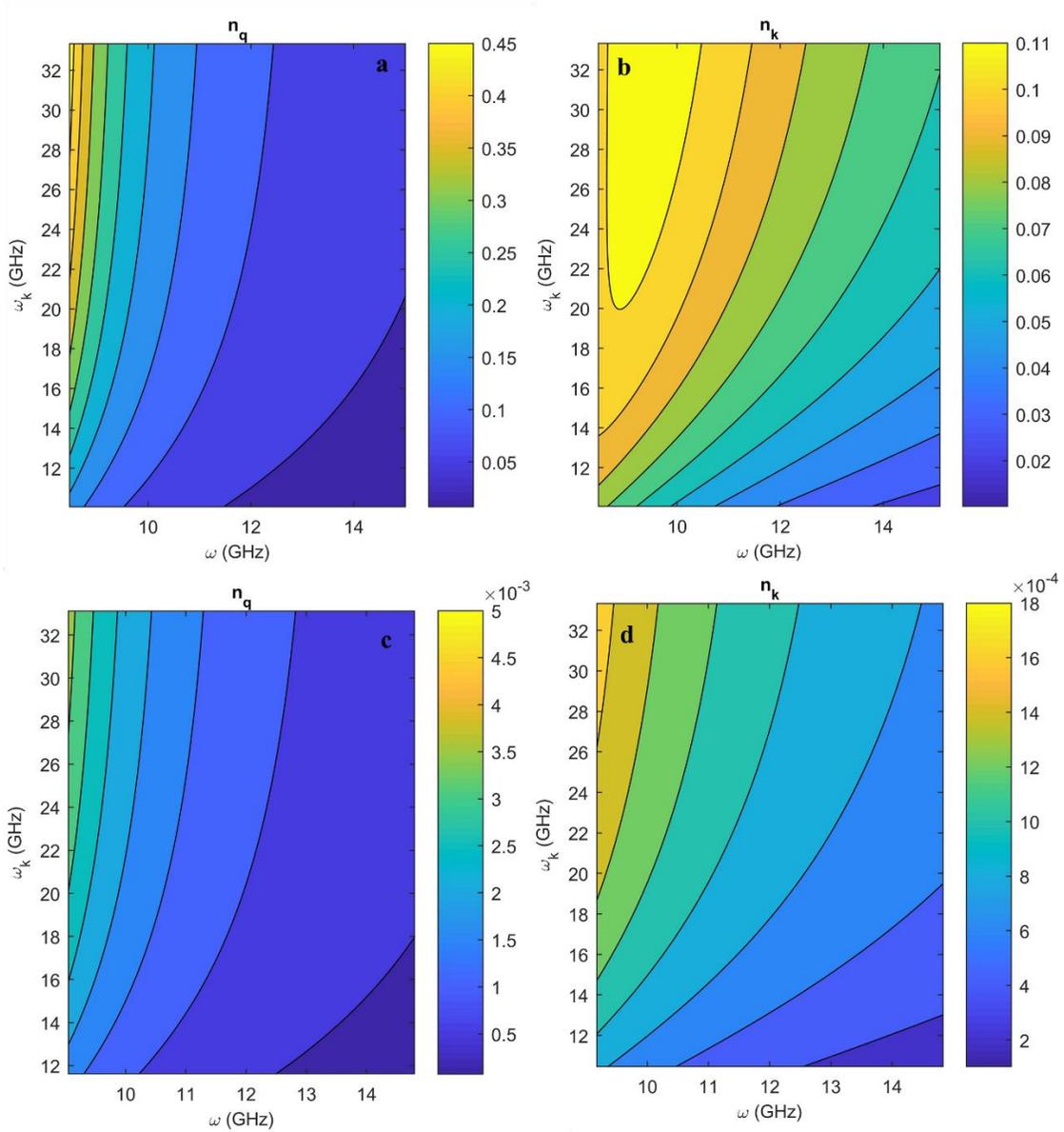

Fig. B1 Effects of the coupling between qubit and reservoir modes on the number of photons of a and c) qubit and b and d) reservoir modes vs. sweeping and reservoir modes frequencies (GHz) at 10 mK; upper row: coupling rate $g_k$ and lower row: coupling rate $0.1g_k$.